\newcommand{\be}{\begin{equation}}
\newcommand{\ee}{\end{equation}}
\documentclass[twocolumn,showpacs,superscriptaddress,amsmath,amssymb]{revtex4}
\usepackage{graphicx}% Include figure files
\usepackage{dcolumn}% Align table columns on decimal point
\usepackage{bm}% bold math

\begin{document}
\title{Rotating solitons supported by a spiral waveguide\\}

\author{Milan S. Petrovi\'c}
\affiliation{Institute of Physics, P.O.Box 57, 11001 Belgrade,
Serbia} \affiliation{Texas A\&M University at Qatar, P.O.Box 23874,
Doha, Qatar}

\author{Aleksandra I. Strini\'c}
\affiliation{Texas A\&M University at Qatar, P.O.Box 23874, Doha,
Qatar} \affiliation{Institute of Physics, University of Belgrade,
P.O.Box 68, 11080 Belgrade, Serbia}

\author{Najdan B. Aleksi\'c}
\affiliation{Texas A\&M University at Qatar, P.O.Box 23874, Doha,
Qatar} \affiliation{Institute of Physics, University of Belgrade,
P.O.Box 68, 11080 Belgrade, Serbia}

\author{Milivoj R. Beli\'c}
\affiliation{Texas A\&M University at Qatar, P.O.Box 23874, Doha,
Qatar}

%\date{\today}
\begin{abstract}
    \noindent We investigate numerically light propagation in a single
spiral waveguide formed in a nonlinear photorefractive medium for a
low spatial frequency of the waveguide rotation. We present the general
procedure for finding solitonic solutions in spiral waveguiding
structures, as well as the variational approach to calculate soliton
parameters analytically. Solitons supported by the spiral waveguide
perform robust stable rotational oscillatory motion, with the
period predicted by their static characteristics, without any
signatures of wave radiation or soliton decay over many rotation periods and
diffraction lengths.
\end{abstract}
%\keywords{Suggested keywords}%Use showkeys class option if keyword

                              %display desired
\maketitle
\section{Introduction}

Nonlinear localized structures or solitons are ubiquitous in nature
\cite{yuri}. Two-dimensional spatial optical solitons are spatially
confined light beams that propagate in nonlinear media, usually along a well-defined propagation direction and with an appropriate transverse profile. However, in general,
the beam propagation need not proceed along a straight line.
Rotating propagation systems provide more interesting dynamics than their
straight counterparts, because the centripetal force modifies the
effect of potentials present and the interaction with the medium or other beams.

Conditions under which two self-guided light beams propagating in a medium with saturating nonlinearity can spiral about
each other in a double-helical orbit are described in \cite{poladian}.
Numerical analysis shows that the three-dimensional propagation and
interaction of mutually incoherent screening spatial solitons in
real (anisotropic) photorefractive crystals typically results in an
initial mutual rotation of their trajectories, followed by damped
oscillations and eventually the fusion of solitons \cite{belic1,belic2}.
Further stabilisation can be achieved by introducing a composite
(rotating "propeller") soliton made of a rotating dipole jointly
trapped with a bell-shaped component \cite{segev2}. Ringlike
localized gain landscapes imprinted in focusing cubic nonlinear
media support stable higher-order vortex solitons \cite{kart1}.

Rotating structures in optical photonic lattices are of special
interest \cite{milan}. Controlled soliton rotation in the optically
induced periodic (Bessel-like) ring lattices is demonstrated in
\cite{wang}. Edge states in an array of evanescently coupled helical
waveguides, arranged in a graphene-like honeycomb lattice, are
responsible for the photonic "topological insulation", in which light
propagates along the edges of a photonic structure, topologically
protected from the scattering off defects \cite{segev1,segev11}. The modulation
causing the topological protection in "photonic graphene" was
achieved by making the waveguides helical; topological protection
cannot be achieved for a wave packet populating a single site
\cite{segev11}. Furthermore, truncated rotating square waveguide
arrays support localized (corner and central) modes that can exist even
in the linear case \cite{kart3}.
Light control in a modulated single-mode waveguide was implemented
for the first time by beam-splitting and adiabatic stabilization of
light in a periodically curved optical waveguide \cite{longhi2}.
Light can be guided in the modulated waveguide because radiation
losses can be successfully suppressed under properly chosen
conditions.

The starting point in understanding these curious optical phenomena is the analogy between paraxial beam
propagation in an optical waveguide with a bent axis and the single-electron dynamics in an atomic system \cite{longhi2}. This analogy originates
from the formal equivalence of the scalar beam propagation equation
for the waveguide in the paraxial approximation and the one-electron
temporal Schr\"{o}dinger equation, represented in the
Kramers-Henneberger (KH) reference frame \cite{henne}. The KH
transformation, originally introduced in atomic physics to
investigate the interaction of a bound electron with superhigh
intensity and high-frequency laser fields, is a transformation to the
moving coordinate frame of the entirely free charged particle
interacting with the applied electromagnetic field. For a high
spatial modulation frequency of the waveguide bending, the refractive
index seen by the beam in the KH frame varies so fast that the beam
dynamics is governed by a cycle-averaged refractive index
potential. Note that if the waveguide has been bent in the (x, z)
plane, the refractive index profile corresponds to a Y splitter
(i.e. the curved waveguide with a short bending period is equivalent
to the Y adiabatic splitter), and the suppression of radiation losses as
well as the wave packet dichotomy are natural consequences of the
appearance of an adiabatic splitter in the cycle-averaged limit
\cite{longhi2}.

The first experimental observation of wave packet
dichotomy and adiabatic stabilization in a periodically bent optical
waveguide was reported in \cite{longhi3}. Adiabatic stabilization
can also be achieved in a three-dimensional waveguide with a helicoidal
(non-planar!) axis bending, and this effect is the optical analogue of the
adiabatic stabilization of a two-dimensional atom in a
high-frequency high-intensity circularly polarized laser field
\cite{longhi1}. In such a way, the cycle-averaged effective waveguide
takes an annular shape, and the launched beam adiabatically evolves
into a ring form \cite{longhi1,garan}. On the other hand, the
three-dimensional spiraling guiding
structures with a shallow refractive index in the approximation of small radius of spiraling, induce a
resonant effect in the form of a coupling and periodic energy exchange
between optical vortices with different topological charges
\cite{kart2}.

In this paper, we investigate numerically light propagation
in the spiraling single waveguide formed in a nonlinear photorefractive medium,
in the limit of a high spatial period (low spatial frequency) of the waveguide rotation and for an
arbitrary helix radius. We present a general procedure for finding
exact fundamental solitonic solutions in the spiraling guiding
structures, based on the modified Petviashvili's iteration method.
We confirm the stability of solitonic solutions by direct numerical
simulation. The existence domain of rotating solitons supported
by a spiral waveguide is relatively wide: below a lower power threshold
they start to radiate and above an upper power threshold they escape
from the waveguide. Solitons supported by the spiral waveguide perform
robust and stable rotary motion over many rotation periods and
diffraction lengths, without any signatures of radiation or decay.
The noise, inevitable in any real physical system, causes a regular
spatial oscillation of the soliton with the period well predicted by
our calculus based on the static characteristics of the soliton. We
also exhibit
an interesting counterintuitive example of beam
spiraling. Finally, we present a variational approach to these phenomena and determine soliton
parameters analytically.

\section{The model}

We start from the well-known paraxial wave equation for the beam
propagation in a nonlinear photorefractive crystal, since it provides
good agreement with experimental data. We choose a photorefractive
medium because there different three-dimensional waveguide structures can
be fabricated easily by use of a femtosecond laser
writing \cite{fabr}. In the steady-state and 3D, the model equation in the dimensionless
computational space (one $x$ or $y$ coordinate unit corresponds to 8.5 $\mu$m and the
$z$ unit corresponds to 4 mm) is given by \cite{prl2005,belic4}:

\begin{equation} \label{PRmodel}
{i \frac{\partial \Psi}{\partial z} + \triangle \Psi + \Gamma
\frac{I+I_w}{1+I+I_w} \Psi = 0}  , \end{equation}

\noindent where $\Psi$ is the beam envelope, $\triangle$ is the
transverse Laplacian, $\Gamma$ is the coupling constant,
$I=|\Psi|^2$ is the laser light intensity measured in units of the
background intensity, and $I_w$ is the optically-induced spiraling waveguide intensity. We
assume that the refractive index change of the waveguide channel has
a radially-symmetric Gaussian shape.

\begin{figure}\vspace{0mm}
\includegraphics[width=50mm]{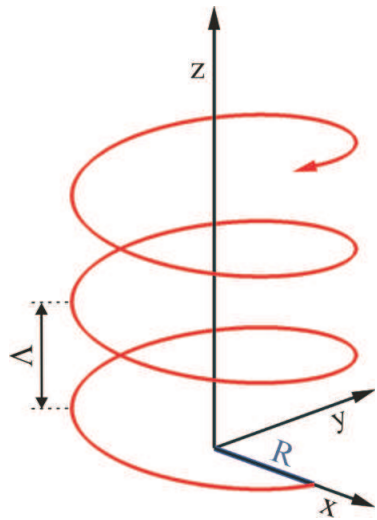}\vspace{0mm}
\caption{\label{Fig1} Sketch of the spiral single waveguide.}
\vspace{0mm}
\end{figure}

The optical waveguide has a helically-twisted axis with sufficiently
small frequency that a guided soliton is adiabatically following
along the waveguide, as it rotates. We transform the coordinates
into a reference frame where the waveguide is straight:

\begin{equation}
{x'=x-R\cos(\Omega z), \quad y'=y-R\sin(\Omega z), \quad z'=z,}
\end{equation}

\noindent where $R$ is the helix radius, $\Omega$ is the spatial frequency
and $\Lambda=2\pi/\Omega$ is the period of rotation. A single spiral
waveguide is sketched in Fig. \ref{Fig1}.

In the transformed reference frame $(x',y')$, the light evolution
is described by:

\begin{equation} \label{prim}
{i \frac{\partial \Psi}{\partial z} = \left( i
\overrightarrow{\nabla}' + \frac{1}{2} \overrightarrow{N}(z')
\right)^2 \Psi - \frac{1}{4} R^2 \Omega^2 \Psi + V \Psi}  ,
\end{equation}

\noindent where $\Psi=\Psi(x',y',z')$ is the transformed envelope, $\overrightarrow{\nabla}' =
\frac{\partial }{\partial x'} \overrightarrow{e_{x'}} +
\frac{\partial }{\partial y'} \overrightarrow{e_{y'}}$ the transformed gradient,
$\overrightarrow{N}(z') = R \Omega [- \sin(\Omega z')
\overrightarrow{e_{x'}} + \cos(\Omega z') \overrightarrow{e_{y'}}
]$ the vector potential, and  $V
= -\Gamma (I+I_w)/({1+I+I_w})$ the scalar potential.
%(tacno je prethodno, prethodni pasus proveren)
After the transformation $\psi(\overrightarrow{r}',z')=\Omega_T
\Psi(\overrightarrow{r}',z')$ \cite{henne}, where
$\overrightarrow{r}' = x' \overrightarrow{e_{x'}} + y'
\overrightarrow{e_{y'}}$ and $\Omega_T = \exp(
\overrightarrow{\delta}\cdot \overrightarrow{\nabla}')$ is a translation
operator for the vector $\overrightarrow{\delta}=- \int_{0}^{z'}
\overrightarrow{N}(\zeta) d \zeta$, Eq. (\ref{prim}) is transformed
in the KH reference frame and can be rewritten in the form:

\begin{equation} \label{primALL}
{i \frac{\partial \psi(\overrightarrow{r}',z')}{\partial z'} = -
 \triangle' \psi(\overrightarrow{r}',z') + V(\overrightarrow{r}'
 + \overrightarrow{\delta}) \psi(\overrightarrow{r}',z')}  .
\end{equation}

\noindent One can see that the new wave function $\psi$ naturally
corresponds to an accelerated frame, and may be
useful in different cases of noninertial frames of reference.

%Gornji pasus - formule su OK

\section{The eigenvalue procedure}

There are no known exact analytical solitonic solutions for our
system. Owing to the symmetry and dynamics of the problem, we are
searching for the self-localized wave packet continuously rotating
and recreating its shape periodically (in every cycle of the
rotation).

%Program solitonska resenja trazi u lokalnom sistemu u kome su
%talasovodi pravi (prim sistem); uzimam i z'=0.

The solitonic solutions can be found from Eq. (\ref{prim}) by
using the modified Petviashvili's iteration method
\cite{petv,yang,najdan}. Our system allows the existence of a
fundamental soliton solution in the form:

\begin{equation} \label{function}
{\Psi(x',y',z')=a(x',y',z') e^{i\mu z'}}
\end{equation}

\noindent where $\mu$ is the propagation constant and $a(x',y',z')$ is a
$z$-periodic complex function with period $\Lambda$. Physical
requirements to obtain a rotationally-invariant solution lead to the
mathematical condition:

\begin{equation} \label{condition}
{\frac{\partial a}{\partial z'} = \Omega y' \frac{\partial
a}{\partial x'} - \Omega x' \frac{\partial a}{\partial y'} }.
\end{equation}

\noindent After substitution of Eqs. (\ref{function}) and
(\ref{condition}) into  Eq. (\ref{prim}), one obtains the soliton equation
in a reference frame where the waveguide is straight:

$$
i \Omega \left[(R\sin(\Omega z')+y')\frac{\partial}{\partial
x'}+(-R\cos(\Omega z')-x')\frac{\partial}{\partial y'} \right]a
$$

\be \label{SolEqPrim} { - \mu a + \triangle' a + \Gamma
\frac{|a|^2+I_w}{1+|a|^2+I_w} a = 0 } . \ee

\noindent Next, without loss of generality, we can calculate soliton
profiles at $z'=0$. Thus, the complex-valued amplitude function
$a(x',y')$, after the separation of linear and nonlinear terms on
different sides of the equation, satisfies a general equation
\begin{equation}
\mathbb{T} - i\Omega R
\frac{\partial a}{\partial y'} - \mu a + \triangle' a + \mathbb{P} a
= \mathbb{Q},
\end{equation} where $\mathbb{T} = i \Omega ( y' \frac{\partial
a}{\partial x'} - x' \frac{\partial a}{\partial y'})$, $\mathbb{P} =
\Gamma {I_w}/({1+|a|^2+I_w})$, and $\mathbb{Q} = - \Gamma
{|a|^2 a}/({1+|a|^2+I_w})$. This equation has to be solved iteratively.

We first perform Fourier transformation of
that equation, to find $\overline{a} = (- \overline{\mathbb{Q}} +
\overline{\mathbb{P}a} + \overline{\mathbb{T}})/(\mu + k_x^2 + k_y^2
+ k_y \Omega R)$, where the overbar denotes the Fourier transform.
Straightforward iteration of this relation unfortunately does not converge,
hence we have to introduce the stabilizing
factors of the form $\alpha = \int [(\mu + k_x^2 + k_y^2 + k_y
\Omega R) \overline{a} - \overline{\mathbb{P}a} -
\overline{\mathbb{T}}] \overline{a}^* d \mathbf{k}$ and $\beta =
-\int \overline{\mathbb{Q}} \overline{a}^* d \mathbf{k}$ into the
equation \cite{najdan}. They do not affect the solution but improve the convergence; this procedure is the essence of the modified Petviashvili's iteration method. The following iteration equation in inverse space is
obtained:

$$
{\overline{a}_{m+1} = \frac{1}{\mu + k_x^2 + k_y^2}}
$$
\be \label{iter} {\times \left[ (\overline{\mathbb{P}a} +
\overline{\mathbb{T}} - k_y \Omega R \overline{a})_m \left(
\frac{\alpha}{\beta} \right)^{\frac{1}{2}}_m -
\overline{\mathbb{Q}}_m \left( \frac{\alpha}{\beta}
\right)^{\frac{3}{2}}_m \right]} , \ee

\noindent where $m$ counts the iterations. In this manner, stable
self-consistent fundamental soliton solutions are found.

\begin{figure}\vspace{2mm}
\includegraphics[width=65mm]{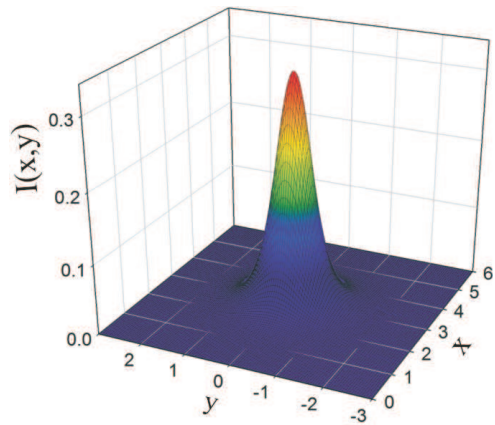}\vspace{0mm}
\caption{\label{Fig2} Fundamental soliton intensity profile for $\mu
= 2 L_D^{-1}$. Parameters: $\Omega$=-0.2 rad$/L_D$, $P_{w}$=0.044,
$R$=3, $\Gamma$=30.} \vspace{0mm}
\end{figure}

The fundamental soliton solution for $\mu = 2 L_D^{-1}$ at $z=0$ is
presented in Fig. \ref{Fig2}. We see that the optical field
intensity $|a(x,y)|^2$ is almost perfectly radially-symmetric. The
main quantity characterizing the spatial soliton is its power $P =
\int\int I dx dy = \int\int |a(x,y)|^2 dx dy$. The imprinted spiral waveguide can
be described in a similar way, because it carries its own intensity, which we take to be Gaussian: $I_w(x',y')
= I_{w0} \exp [-(x'^2+y'^2)/W_w^2]$; this gives $P_w = \int\int
I_w(x',y') dx' dy' = \pi I_{w0} W_w^2$.

\begin{figure}\vspace{0mm}
\includegraphics[width=75mm]{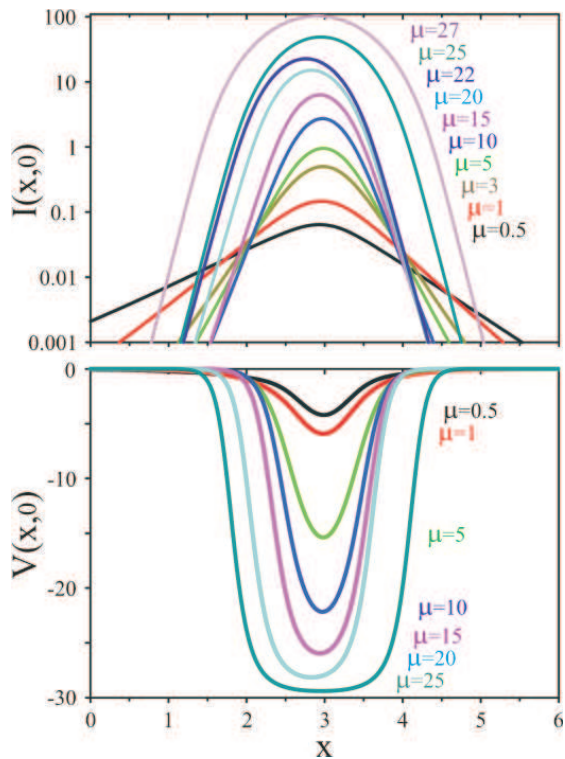}\vspace{0mm}
\caption{\label{Fig3} Fundamental soliton intensity profiles (top)
and the corresponding potential profiles (bottom) at y = 0.
Parameters are as in Fig. \ref{Fig2}, except $P_{w}$=0.05
($I_{w0}=0.1$, $W_w=0.4$). } \vspace{0mm}
\end{figure}

In this manner, one finds a family of fundamental solitonic solutions with
different propagation constants and beam powers (Fig. \ref{Fig3})
for each set of reasonable physical parameters. It is interesting to note
that the existence domain is rather wide, {\it i.e.} although soliton
widths are similar, their intensities and the corresponding potentials
differ significantly from one another. The solutions
are located close to the waveguide center (the helix radius $R=3$ here)
and exactly at the potential barrier minimum, as expected. Because
of the saturation nature of photorefractive nonlinearity, for large
intensities the potential $V$ tends to $-\Gamma_+$ ($\Gamma=30$ here).

\begin{figure}\vspace{0mm}
\includegraphics[width=75mm]{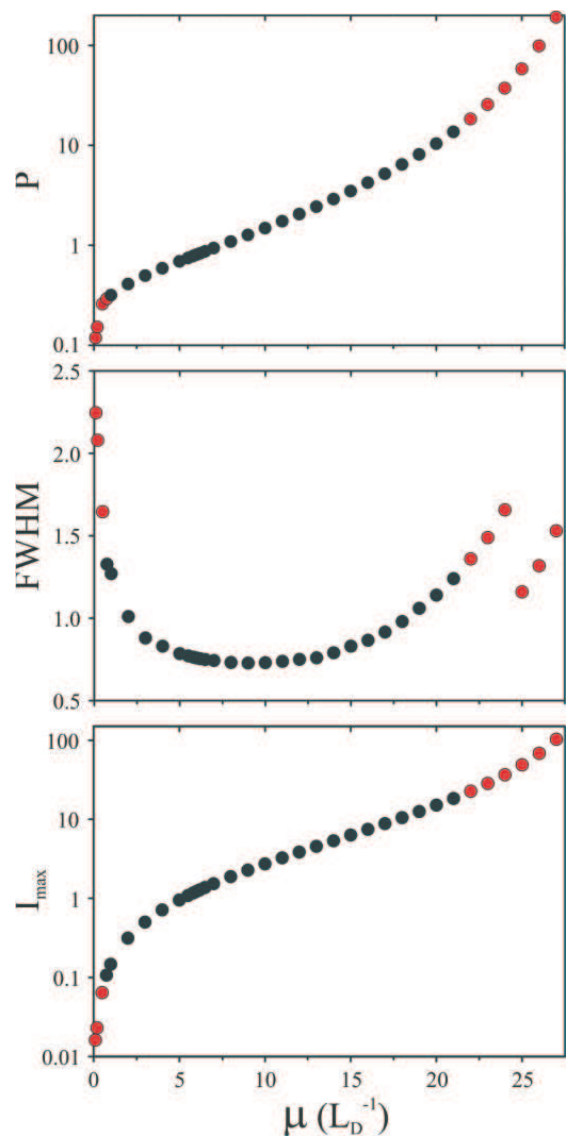}\vspace{0mm}
\caption{\label{Fig4} Fundamental soliton power, width, and peak
intensity as functions of the propagation constant. Black dots
represent the stable rotary solitonic solutions, red dots the
unstable solutions. Parameters are as in Fig. \ref{Fig3}.}
\vspace{0mm}
\end{figure}

The most important fundamental soliton charcteristics (the soliton
power, width, and peak intensity) as functions of the propagation
constant are shown in Fig. \ref{Fig4}. We see that the
soliton width has a minimum value; this region corresponds to very
stable rotating soliton propagation. One can notice from Fig.
\ref{Fig4} that the obtained solitonic solution is stable, according
to the Vakhitov-Kolokolov stability criterion \cite{vk}, which claims
that the solitary wave should be stable as long as $dP/d\mu>0$. We
marked the unstable solutions in Fig. \ref{Fig4} in red: below the lower power
threshold they start to radiate, and above the upper power threshold
they escape from the potential well.

\section{Soliton dynamics}

In order to track the trajectory of solitons during propagation
through the (nonlinear) photorefractive crystal, it is
necessary to look inside the crystal volume, which experimentally is
hardly feasible, because the solitons hardly radiate and are invisible from the side. Therefore, numerical simulation is preferred. Numerical
procedure applied to the propagation equation is the split-step beam
propagation method based on the fast Fourier transform (fourth-order
symplectic algorithm). We launch a soliton (from the point $y=0$) with an initial angular
momentum in the form of an input phase tilt in the  $y$ direction, which
introduces beam velocity tangential to the spiral waveguide;
the helix radius is constant here, in difference to \cite{longhi1}. The beam
can be set into a steady spiraling motion with a period dictated by the
period of the helical waveguide.

\begin{figure}\vspace{0mm}
\includegraphics[width=\columnwidth]{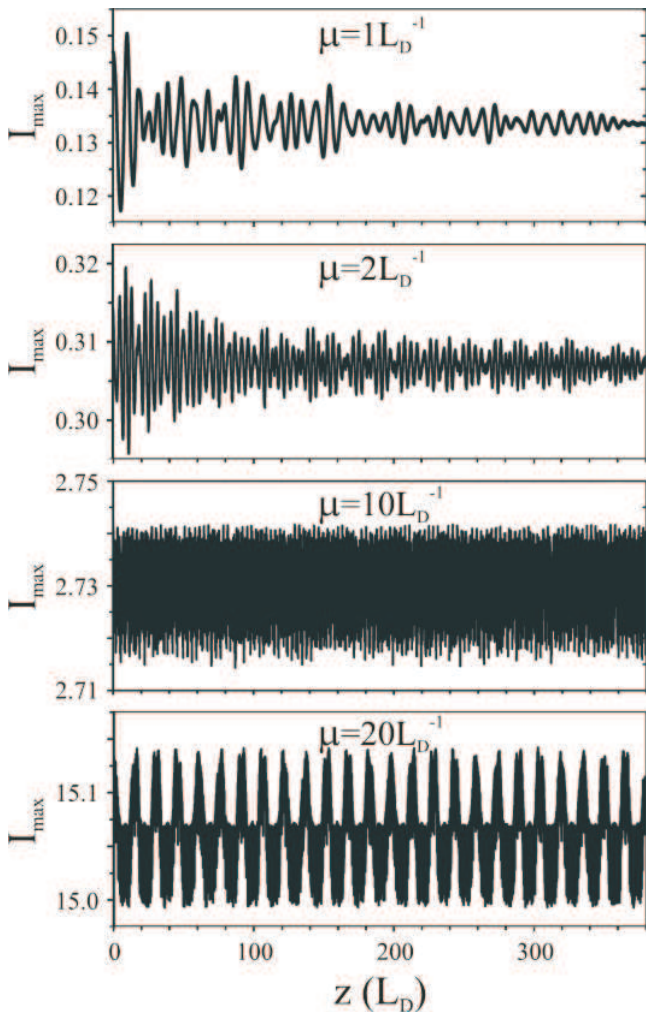}\vspace{0mm}
\caption{\label{Fig5} Peak intensity as a function of the
propagation distance for several different values of the propagation
constant. Parameters are as in Fig. \ref{Fig3}.} \vspace{0mm}
\end{figure}

To check the iterative procedure for finding solitons, we
propagate this input solution (in the stationary frame of reference): the
peak intensity as a function of the propagation distance for several
different values of the propagation constant is shown in Fig.
\ref{Fig5}. The existence domain of the rotating solitons supported
by the spiral waveguide is rather wide (as presented in Fig.
\ref{Fig4}). At the lower power threshold (the case $\mu= 1 L_D^{-1}$ from
Fig. \ref{Fig5}) the solitons radiate energy in the beginning. In the
central part of the existence domain the fundamental solutions perform
persistent stable rotary motion (the three remaining cases from Fig.
\ref{Fig5}), while the power conservation is almost perfect (although
the peaks slightly oscillate). Above the upper power threshold, the
solitons escape from the waveguide. On the other hand, when the
rotation frequency exceeds a critical value, no localized modes can
be found, since the potential barrier cannot produce the required
centripetal force at such high frequencies.

The trajectory of the light beam is defined as the spatial
expectation value of its transverse coordinates, weighted by the beam
intensity:

\begin{equation} \label{x exp}
{\langle x \rangle(z) = \frac{1}{P} \int_{-\infty}^{\infty} dy
\int_{-\infty}^{\infty} x |\Psi(x,y,z)|^2 dx} , \quad
\end{equation}

\begin{equation} \label{y exp}
{\langle y \rangle(z) = \frac{1}{P} \int_{-\infty}^{\infty} dx
\int_{-\infty}^{\infty} y |\Psi(x,y,z)|^2 dy} .
\end{equation}

\begin{figure}\vspace{0mm}
\includegraphics[width=70mm]{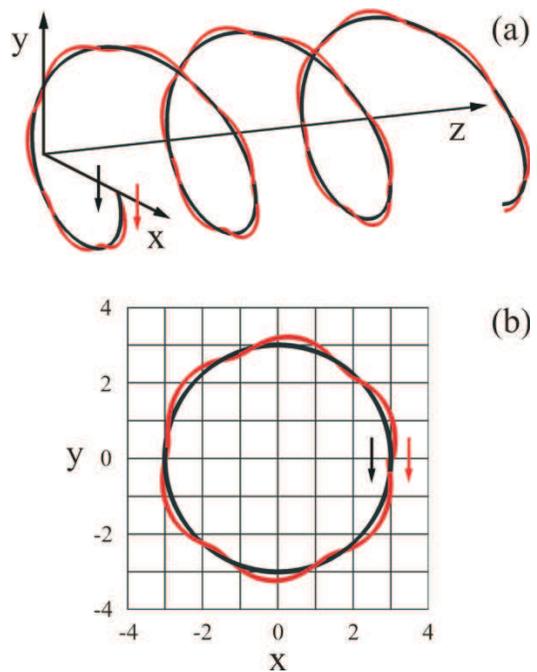}\vspace{0mm}
\caption{\label{Fig6} Typical trajectory of a rotating soliton (red
line) supported by the spiral waveguide (black line): The fundamental
soliton oscillates regularly around the waveguide during propagation.
(a) 3D view for the propagation distance $z = 100 L_D = 3.183 \Lambda$.
(b) 2D view for $z = \Lambda$. Parameters are as in Fig. \ref{Fig3},
$\mu = 15 L_D^{-1}$.} \vspace{0mm}
\end{figure}

\noindent A characteristic oscillatory trajectory of the rotating soliton
supported by the spiral waveguide is presented in Fig. \ref{Fig6}.
Unavoidable non-ideal beam launching, which introduces numerical
uncertainty in our system, causes a spatial oscillation of the soliton
that propagates stably around the waveguide and in this way demonstrates a
novel interesting type of soliton dynamics.

From Fig. \ref{Fig6}(b),
which covers one full oscillation period of the helical waveguide,
one can notice that the fundamental soliton oscillates very regularly
around the waveguide during propagation. We should also mention that, opposite
to the breather solitons in nonlocal media where the oscillation appears
in the amplitude as well as in other soliton parameters
\cite{najdan}, here we have spatial oscillation with practically
constant amplitude (see Fig. \ref{Fig5}). Although our system is not
integrable, rotating solitons supported by the spiral waveguide move
self-consistently as particles in a potential created by the induced
change in the refractive index \cite{belic3}.

\begin{figure}\vspace{0mm}
\includegraphics[width=\columnwidth]{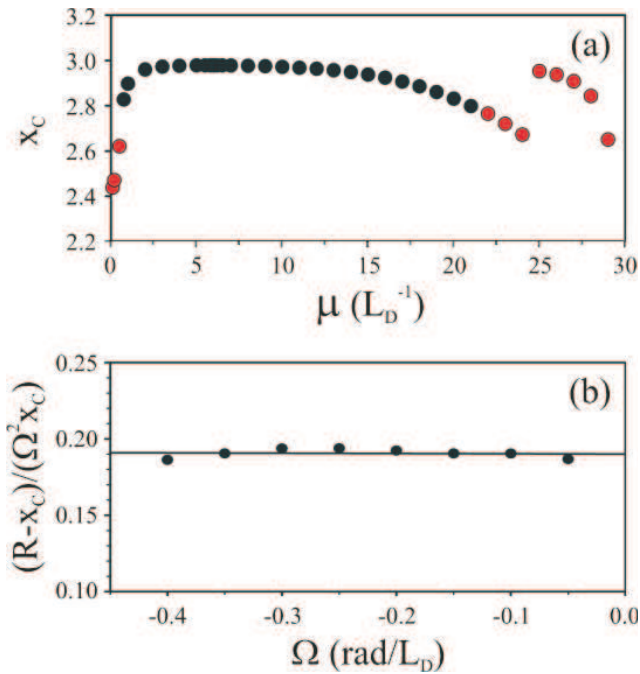}\vspace{0mm}
\caption{\label{Fig7} (a) The initial soliton center position $x_c$
as a function of the propagation constant $\mu$. Black dots
represent the stable rotary solitonic solutions, red dots the
unstable solutions. Parameters are as in Fig. \ref{Fig3}. (b) The
quantity $(R-x_c)/(\Omega^2 x_c)$ is practically independent of the
frequency of rotation $\Omega$. Parameters are as in Fig.
\ref{Fig3}, except for $P_{w}$=0.1 and $P$=0.5.} \vspace{0mm}
\end{figure}

An interesting question to ponder is, is it possible to connect static with dynamic characteristics of
rotating solitons? The short answer is yes. Let's consider the initial
soliton center position $x_c$, the quantity obtained in the eigenvalue
procedure, shown in Fig. \ref{Fig7}(a) as function of the
propagation constant $\mu$. We see that $x_c < R$ always; the
difference is the smallest in the domain of parameter space where
the solitonic solutions are stable. Because rotating solitons behave as
particles, the centripetal force acting on the beam during uniform
circular motion is of the form $m \Omega^2 x_c$, where $m$ is the
soliton "mass" (proportional to the soliton power $P$). On the other
hand, there is a force associated with the potential created by the
waveguide; in the first approximation we can consider that the force is
proportional to the distance from the equilibrium (the dynamical elongation), {\it i.e.} of
the form $k (R - x_c)$ where $k$ is the "force constant". The two forces
are equal, and if our assumption is correct, the next
relation should be constant:

\begin{equation} \label{const}
{\frac{m}{k} = \frac{R-x_c}{\Omega^2 x_c} = \textrm{const}} ,
\end{equation}

\noindent for each $\Omega$ and the constant values of both $P$ and
$P_w$ (it is easy to understand that if $P = \textrm{const}$ and
$P_w = \textrm{const}$, then also both soliton "mass" and "force
constant" are constant). From Fig. \ref{Fig7}(b) is clear that
quantity $(R-x_c)/(\Omega^2 x_c)$ is independent of the frequency of
rotation, and Eq. (\ref{const}) is fulfilled.

Now when we know the nature of interaction in our system, we can
understand dynamics of oscillating spatial solitons better. Owing to
the analogy with stretched spring, harmonically oscillating solution
is expected for sufficient small oscillation amplitudes, with the period

\begin{equation} \label{period}
{T = \frac{2 \pi}{\Omega}  \sqrt{ \frac{R-x_c}{x_c}} } .
\end{equation}

\begin{figure}\vspace{0mm}
\includegraphics[width=\columnwidth]{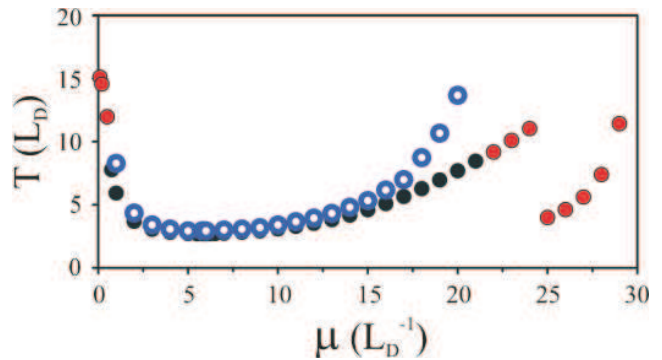}\vspace{0mm}
\caption{\label{Fig8} The period of small oscillations $T$ as a
function of the propagation constant. For the stable (black dots)
and unstable (red dots) rotating solitons period is given by Eq.
(\ref{period}); blue dots represent results obtained in numerical
simulations. Parameters are as in Fig. \ref{Fig3}.} \vspace{0mm}
\end{figure}

\noindent The period of small oscillations $T$ as a function of the
propagation constant is represented in Fig. \ref{Fig8}. To check
this result, we propagate a fundamental soliton in the case of small
perturbations, and find that it oscillates regularly with a period
in good agreement with that obtained by the Eq. (\ref{period}). In
such a way, main characteristics of dynamical behavior of system can
be predicted from the eigenvalue procedure.

\begin{figure}\vspace{0mm}
    \includegraphics[width=70mm]{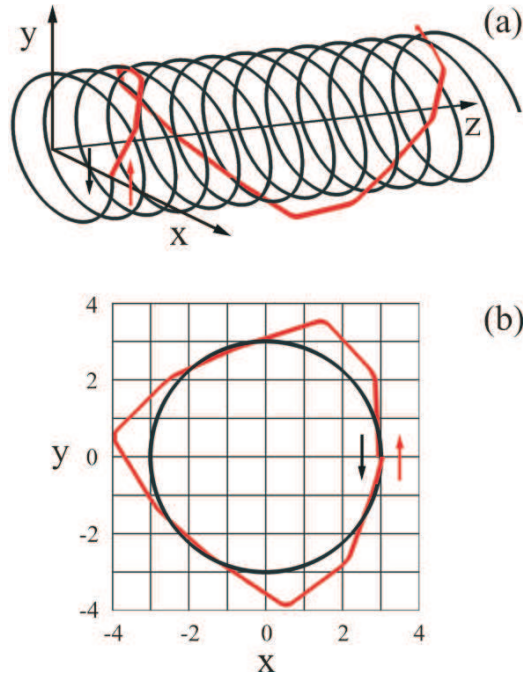}\vspace{0mm}
    \caption{\label{Fig9} Untypical trajectory of rotating soliton (red
        line) supported by spiral waveguide (black line): fundamental
        soliton is launched counterclockwise and rotates in this direction
        during propagation, while waveguide spirals in opposite direction.
        (a) 3D view for propagation distance $z = 400 L_D$. (b) 2D view for
        $z = 294 L_D$ (one period). Spiral waveguide channel has a
        hyper-Gaussian shape with $P_{w}$=0.044, $\mu = 26 L_D^{-1}$ and
        $P$=98.9; other parameters are as in Fig. \ref{Fig3}.} \vspace{0mm}
\end{figure}

Spiraling single waveguide supports rich light beam dynamics.
Soliton from upper power threshold, launched with a carefully chosen
initial angular momentum in the direction opposite to the waveguide
direction of spiraling, instead to escape from the potential
barrier, becomes attracted by the spiral waveguide periodically,
rapidly changing its orbit, and can be set into rotation. Such an
example of untypical trajectory of rotating soliton supported by
spiral waveguide is shown in Fig. \ref{Fig9}: fundamental soliton
rotates counterclockwise during propagation, while the waveguide spirals
in the opposite direction.

\section{VARIATIONAL APPROACH}

Here, we introduce the concept and the most important ideas of
the variational approach (VA) to helical waveguides and illustrate the main
challenges and problems; the complete VA procedure for "slow"
helical waveguiding in media with arbitrary nonlinearity will be
presented elsewhere.

In order to better understand dynamical phenomena concerning beam
propagation along a helical waveguide, we apply a powerful approximate
technique to the governing Eq. (\ref{PRmodel}), based on the VA. The
key idea is to decouple nonlinearity from the waveguide

\begin{equation} \label{VAprim}
{i \frac{\partial \Psi}{\partial z} + \triangle \Psi + \Gamma
I_w(x',y') \Psi + \Gamma \frac{I}{1+I} \Psi = 0}  , \end{equation}

\noindent which is justified in the shallow waveguide approximation
(small $I_w$). After the transformation to the moving coordinate frame,
in the Lagrangian density formalism we assume a Gaussian beam
solution whose parameters vary along z, and make an ansatz:

\begin{widetext}
\begin{equation} \label{VA_ansatz}
{ \Psi = A \exp \left[-{\frac{(x'-X_C)^2}{2W^2}} -
{\frac{(y'-Y_C)^2}{2W^2}} + i C_x (x'-X_C)^2 + i C_y (y'-Y_C)^2 + i
S_x (x'-X_C) + i S_y (y'-Y_C) + i \varphi \right]} ,
\end{equation}
\end{widetext}

% ista formula samo podeljena: NE brisati
%\begin{align} \notag
%\Psi = A &\exp \left[-{\frac{(x'-X_C)^2}{2W^2}} -
%{\frac{(y'-Y_C)^2}{2W^2}} \right]\\
%\times &\exp \left[i C_x (x'-X_C)^2 + i C_y (y'-Y_C)^2 \right] \notag\\
%\times &\exp \left[ i S_x (x'-X_C) + i S_y (y'-Y_C) + i \varphi
%\right] ,  \label{VA_ansatz}
%\end{align}

\noindent where $A$ is the amplitude, $W$ is the width of the beam,
($X_C$, $Y_C$) is the transverse position of the beam's center, $C_x$
and $C_y$ are the wave front curvatures along $x$ and $y$, $S_x$ and $S_y$ are drift
"velocity" components, and $\varphi$ is the nonlinear phase shift.
In the first approximation we analyze the dynamics of an axially
symmetric beam in a symmetrical waveguide. In the optimization
procedure, the first variation of the corresponding functional must
vanish, if trial functions are chosen properly. The dynamics of the
beam is described by the motion of a representative particle in the
four dimensional nonstationary potential. The analysis is fairly
complex, and after a lot of algebra, one obtains a nonlinear equation
for beam width $W$

\begin{align} \notag
&\frac{1}{W^2} - \frac{\Gamma I_{w0} W_w^2 W^2
(W_w^2+W^2-Q^2)}{(W_w^2+W^2)^3}\\
&\times  \exp \left(-{\frac{Q^2}{W_w^2+W^2}} \right) - \Gamma D = 0
, \label{VA_zero}
\end{align}

%\begin{equation} \label{VA_zero}
%{ \frac{\Gamma I_{w0} W_w^2}{(W_w^2+W^2)^2} \exp
%\left(-{\frac{Q^2}{W_w^2+W^2}} \right)-\frac{R \Omega^2}{4 Q} = 0} ,
%\end{equation}

\noindent where

\begin{align} \notag
&Q \equiv \sqrt{X_C^2 + Y_C^2} = 2 \frac{W_w^2+W^2}{R \Omega^2 W^2}\\
&\times \left(\Gamma D - \frac{1}{W^2} + \sqrt{\left(\Gamma D -
\frac{1}{W^2}\right)^2 +  \frac{R^2 \Omega^4 W^4}{4(W_w^2+W^2 )}}
\right) , \label{VA_Q}
\end{align}

\noindent is the transverse distance of the beam's center, and

\begin{equation} \label{VA_D}
{ D = -{\frac{\ln(1+A^2)+Li_2(-A^2)}{A^2}} }
\end{equation} is related to the beam's amplitude. The integral $Li_2(\zeta)$
defined by $Li_2(\zeta)=\int_{\zeta}^0 dt \ln(1-t)/t$ is the
dilogarithm function. Equation (\ref{VA_zero}) may be regarded as a
procedure to find zeros numerically; for the given value of $A$ one
calculates $Li_2$ and $D$ first, and after that the zeros of Eq.
(\ref{VA_zero}) can be found easily. In fact, there are two zeros,
but only the lower one is stable.

\begin{figure}\vspace{0mm}
    \includegraphics[width=75mm]{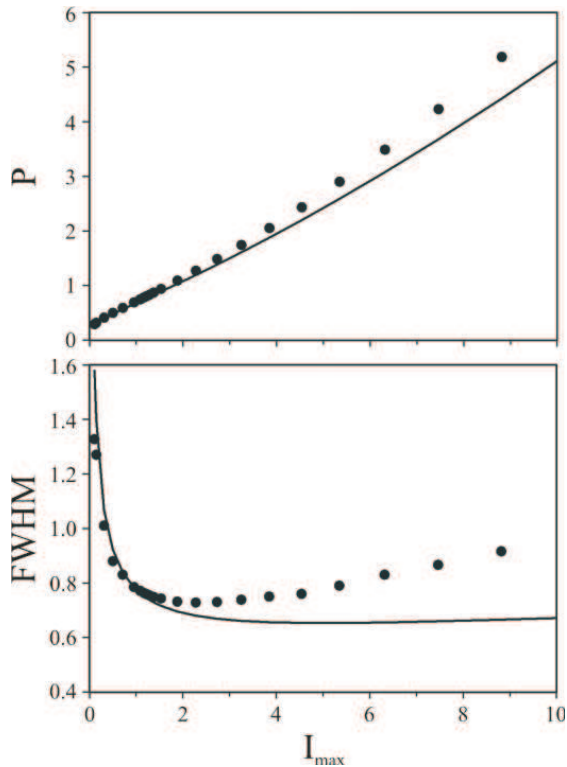}\vspace{0mm}
    \caption{\label{Fig10} Fundamental soliton power and width as
        functions of the peak intensity. Black dots represent stable rotary
        solitonic solutions obtained numerically, black solid line
        represents the results of the variational approach. Parameters are
        as in Fig. \ref{Fig3}.} \vspace{0mm}
\end{figure}

In the dynamical case, the steady state does not exist, but one can still estimate
soliton parameters. The question is, how good they are? The soliton
power $P = \pi A^2 W^2$ and the width FWHM=2$\sqrt{\ln(2)}$ $W$, as
functions of the peak intensity, are shown in Fig. \ref{Fig10}. We note good agreement between the results of
variational approach and the numerical solitonic solutions for the
smaller (physically more acceptable) values of peak intensity, for which the VA is valid. In
this region (peak intensity between 0.1 and 7 for the given set
of physical parameters, and $0.75\leq\mu\leq15$) where the soliton
in its initial position closely overlaps the waveguide (see Fig.
\ref{Fig7}(a)), the period of small oscillations $T$ is also
calculated in a satisfactory manner (see Fig. \ref{Fig8}). This
means that the VA cannot be applied to cases of large displacement
between the beam and the waveguide peak position in the equilibrium state,
because Eq. (\ref{VAprim}) must be fulfilled at each transverse
point.

\section{Conclusions}

In this paper, we have studied numerically nonlinear light propagation in a helically twisted
optical waveguide formed in a photorefractive medium. We have presented a general procedure for finding exact
fundamental solitonic solutions in the spiraling guiding structures,
based on the modified Petviashvili's iteration method. A region in the parameter space is determined, in which stable rotating solitons exist. Below the lower power threshold, the rotating
solitons supported by the spiral waveguide
start to radiate, and above the upper threshold they escape from the waveguide. Their
stability was confirmed by direct numerical simulations. Spiralling
spatial solitons supported by the 3D helical waveguide structure perform
robust and stable rotational-oscillatory motion, without any
signatures of radiation or decay, over many rotation periods and
diffraction lengths. Inevitable numerical inaccuracy causes a regular spatial oscillation of
the soliton, with the period well predicted by our calculated value.

We have
developed a variational approach to find an approximate Gaussian beam solution and used it to calculate soliton parameters
analytically. The "slow" helical waveguiding can be considered as a
kind of dynamic localization, because the localized particle (the soliton
here) periodically returns to its initial state, following the
periodic change of the driving field (the helical waveguide). The spiraling
single waveguide provides an excellent opportunity for studying
phenomena of light propagation balanced between discreteness and
nonlinearity.

\vspace{0pt}
\begin{acknowledgments}
This work was supported by the Ministry of Science of the Republic
of Serbia under the projects OI 171033, 171006, and by the NPRP
7-665-1-125 project of the Qatar National Research Fund (a member of
the Qatar Foundation). Authors acknowledge supercomputer time
provided by the IT Research Computing group of Texas A\&M University
at Qatar. MRB acknowledges support by the Al Sraiya Holding Group.
\end{acknowledgments}

\end{document}